\documentstyle[twoside,fleqn,espcrc2,epsfig]{article}
\title{ SW action for the lattice Schwinger 
model\thanks{Presented by C. Hoelbling; E-mail: hch@bu.edu}}

\author{Ch.~Hoelbling\address{Department of Physics, Boston
University, Boston, MA 02215, USA}, C.~B. Lang\address{Institut f\"ur
theoretische Physik, Universit\"at Graz, A-8010 Graz, Austria} and
R.~Teppner$^\textrm{\scriptsize b}$}

\begin{document}
\pagestyle{empty}

\begin{abstract}
We study some aspects of the ${\cal O}(a)$ improved
Sheikholeslami-Wohlert (SW) action for the lattice Schwinger model. 
We find some improvement concerning the distribution of eigenvalues 
of the Dirac operator and of the masses but little or no
improvement for rotational invariance of correlators or dispersion relations.
\end{abstract}

\maketitle

\section{INTRODUCTION}

The ${\cal O}(a)$ improved action, as introduced by Symanzik
\cite{Sy83} and  proposed for lattices fermions by Sheikholeslami and
Wohlert \cite{ShWo85} has become a popular and relatively simple tool
for improving lattice predictions over the last years. It is therefore
interesting to investigate how different observables are affected by
the ${\cal O}(a)$ improvement of the standard Wilson action in a simple
and well-known toy model.

Our testing ground is the 2-flavor Schwinger model (QED in 2D), which
is well-known analytically and quite undemanding computationally. We
perform a full fermion HMC simulation and study bosonic masses and
dispersion relations, rotational invariance and the eigenvalue spectrum
of the action.

The SW action is obtained from the naive fermion action by a rotation of the
fermionic fields
\begin{equation}
\label{frot}
\Psi\rightarrow(1-\frac{1}{2}\,{\slash\!\!\!\!D})\,\Psi\quad ,\qquad
\bar{\Psi}\rightarrow\bar{\Psi}\,(1+\frac{1}{2}\,\stackrel{\leftarrow}
{\slash\!\!\!\!D}) 
\end{equation}
and discarding the ${\cal O}(a^2)$ terms in the resulting
action.

This results in the Wilson action plus an additional SW term
\begin{equation}
 S_\textrm{\scriptsize SW}=-\kappa\,c_\textrm{\scriptsize SW}\,
\frac{i}{2}\,\sum_x F_{\mu\nu} \bar{\Psi}_x\sigma_{\mu\nu}\Psi_x
 \; .
\end{equation}
The coefficient $c_\textrm{\scriptsize SW}$ is $1$ for tree-level
improvement and has to be determined perturbatively or
non-perturbatively in 4D. In our case of the superrenormalizable
Schwinger model, the coupling $e$ has dimension $1/a$, so that in a
perturbative expansion, where the next-to-leading order graph has two
vertices, $c_\textrm{\scriptsize SW}=1+{\cal O}(a^2e^2)$. Therefore for
an ${\cal O}(a)$ improvement $c_\textrm{\scriptsize SW}=1$.

\section{RESULTS}

\subsection{Phase diagram}

\begin{figure}[t]
\vspace*{-1mm}
\begin{center}
\epsfig{file=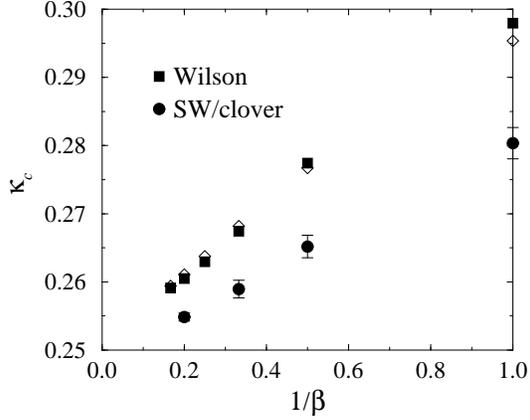,width=7cm}
\end{center}
\caption{\label{phd}Phase diagram for Wilson and clover action on an
$8\times 8$ lattice. The diamonds indicate the results (Wilson action)
of an extrapolation to infinite lattice volume from results at various
lattice sizes \cite{HiLaTe98}.}
\end{figure}

First we looked at the phase diagram comparing Wilson to SW action
(Fig. \ref{phd}). We determined $\kappa_c$ at a given $\beta$ by PCAC
methods, i.e. measuring an observable, which is proportional to the
effective fermion mass at five different values of $\kappa$ and 
finding $\kappa_c$ by interpolation (for more details cf. 
\cite{HiLaTe98}).

As expected, the critical line moves closer to the continuum critical
value of $\beta_c=0.25$. However, we find that for both actions the
leading correction goes like
\begin{equation}
\kappa_c=0.25+ {\cal O}(1/\beta)=0.25+{\cal O}(a^2) \;.
\end{equation}
We found the volume dependence of $\kappa_c$ to be weak (see \cite{HiLaTe98}
for a discussion for the Wilson action).

\subsection{Spectrum}

We also determined the eigenvalue spectrum of the fermion matrix for different
configurations.

As has been pointed out recently, the fixed point action
\cite{HaNi98,FaLaWo98} as well as the overlap action \cite{Ne98} have
an eigenvalue spectrum of circular shape; this is related to the fact
that they satisfy the Ginsparg-Wilson condition \cite{GiWi82} and
therefore have good chiral properties.

\begin{figure}[t]
\begin{center}
\epsfig{file=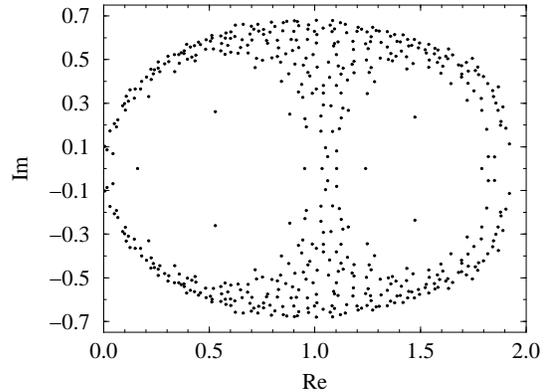,width=7cm}
\end{center}
\caption{\label{spec}The eigenvalue spectrum of the SW Dirac operator for a
typical configuration on a $16\times 16$ lattice at $\beta=2$}
\end{figure}

As can be seen in Fig.\ref{spec}, the SW spectrum differs from the
Wilson spectrum (see e.g.\cite{GaHiLa97a}) but does not improve
much towards a circular shape. The change of the spectrum is consistent
with the observations made in \cite{GaHi98} for 4D SU(2) theory.

\subsection{Rotational invariance}

Another interesting aspect is, whether one may observe any
improvement in the rotational invariance of propagators. For this
purpose, we measured
\begin{equation}\label{rotprop}
\langle\bar{u}(0)\,\sigma_3\,d(0)\,\bar{d}(x)\,\sigma_3\,u(x)\rangle \quad ,
\end{equation}
for all distances across the lattice and plotted the correlation
function vs. $|x|$ (where $u$ and $d$ denote the two fermion flavors).
For a comparison, this was done at the same $\beta$
and for the same effective fermion mass for both actions. For the SW
situation the fermion fields were also improved according to
(\ref{frot}).  Although the propagator is an off-shell quantity one
might expect some improvement, since we also improved the fermion
fields.

\begin{figure}[t]
\begin{center}
\epsfig{file=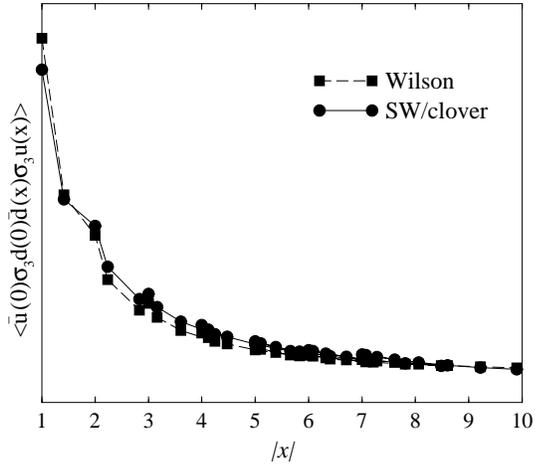,width=7cm}
\end{center}
\caption{\label{rot}Plot of the correlation function (\ref{rotprop}) 
as measured at $\beta=2$ on
a $16\times 16$ lattice at the respective $\kappa_c$. Using the clover
action leads to no improvement in the rotational invariance.}
\end{figure}

As can be seen from Fig. \ref{rot} there is no noticeable improvement in
the rotational invariance for short distances.  This is our first
indication, that also high momentum observables are not improved.

\subsection{Dispersion relations}

The spectrum of the massless $N$-flavor Schwinger model contains one free
massive (isosinglet-vector) meson with mass $m_S=\sqrt{N/\beta\pi}$ and
$N^2-1$ free massless (isomultiplet-vector) mesons.

\begin{figure}[t]
\begin{center}
\epsfig{file=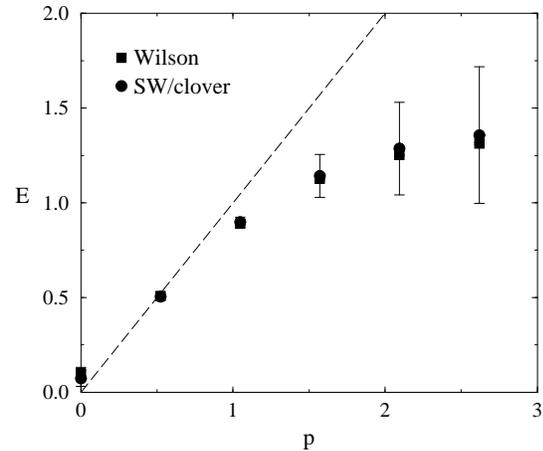,width=7cm}
\end{center}
\caption{\label{disp0}Dispersion relation for the massless vector boson
at $\beta=2$ on a $12^2$ lattice. We find no noticeable no improvement 
towards the continuum shape (dashed line) for higher momentum modes.}
\end{figure}

We measured the dispersion relation of both the massive and the
massless bosons and found -- compared to the results for the
Wilson action -- no improvement (Fig.\ref{disp0}). The higher momentum
modes are clearly deviating from the continuum behavior. For
the fixed point action drastically better behavior has been observed
\cite{FaLaWo98}.

For the massive boson one expects
improved scaling; indeed there seems to be evidence that the mass
estimate at small $\beta$ is improved with the SW action.
However, the statistics of our data does not allow conclusive
statements at this point.

\section{CONCLUSIONS}

We studied the effect of ${\cal O}(a)$ SW-improvement of the Wilson
action in the unquenched $2$-flavor lattice Schwinger model.

We found that the critical value of the hopping parameter $\kappa_c$
moves closer to its continuum value $0.25$. We observed a change in the
eigenvalue spectrum of the Dirac operator but we found no significant
improvement towards e.g. a circular shape. We found no improvement in
the rotational symmetry of correlation functions or in the dispersion
relations at higher momentum.

Since it is well established from QCD simulations that meson masses
are improved by using the SW action, we conclude that the ${\cal
O}(a)$ SW-improvement works best for low momentum states and that there
might be no relevant improvement for observables connected with higher
momentum states.

This work was supported by Fonds 
zur F\"orderung der Wissen\-schaft\-lichen Forschung 
in \"Osterreich, Project P11502-PHY.


\begin{thebibliography}{10}
\bibitem{Sy83}
K. Symanzik,
\newblock Nucl. Phys. B 226 (1983) 187.

\bibitem{ShWo85}
B. Sheikholeslami and R. Wohlert,
\newblock Nucl. Phys. B. 259 (1985) 572.

\bibitem{HiLaTe98}
I. Hip, C.~B. Lang and R. Teppner,
\newblock Nucl. Phys. (Proc. Suppl.) 63 (1998) 682,
and in preparation.

\bibitem{HaNi98}
P. Hasenfratz,
\newblock Nucl. Phys.B (Proc. Suppl.) 63A-C (1997) 53;
P. Hasenfratz, V. Laliena, and F. Niedermayer,
\newblock Phys. Lett. B 427 (1998) 125.

\bibitem{FaLaWo98}
F. Farchioni, C.~B. Lang, and M. Wohlgenannt,
\newblock Phys. Lett. B 433 (1998) 377;
cf. also I. Hip, F. Farchioni, C.~B. Lang, and M. Wohlgenannt,
hep-lat/9809049, contrib. to this conference.

\bibitem{Ne98}
H. Neuberger,
\newblock Phys. Lett. B 417 (1998) 141;
Phys. Lett. B 427 (1998) 353. 

\bibitem{GiWi82}
P.~H. Ginsparg and K.~G. Wilson
\newblock Phys. Rev. D25 (1982) 2649.


\bibitem{GaHiLa97a}
C.~R. Gattringer, I. Hip, and C.~B. Lang,
\newblock Nucl. Phys. B 508 (1997) 329.

\bibitem{GaHi98}
C.~R. Gattringer and I. Hip,
\newblock hep-lat/9806032,
to appear in Nucl. Phys. B.


\end{thebibliography}
\end{document}